# Soliton mobility in nonlocal optical lattices

Zhiyong Xu, Yaroslav V. Kartashov,* and Lluis Torner

*ICFO-Institut de Ciencies Fotoniques, and Universitat Politecnica de Catalunya,*

*Barcelona, Spain*

We address the impact of nonlocality in the physical features exhibited by solitons supported by Kerr-type nonlinear media with an imprinted optical lattice. We discover that nonlocality of nonlinear response can profoundly affect the soliton mobility, hence all the related phenomena. Such behavior manifests itself in significant reductions of the Peierls-Nabarro potential with increase of the degree of nonlocality, a result that opens the rare possibility in nature of almost radiationless propagation of highly localized solitons across the lattice.

*PACS numbers: 42.65.Tg, 42.65.Jx, 42.65.Wi*

During the last two decades spatial solitons have become a subject of intense investigation because of their unique physical features [1]. Properties of solitons supported by media with local nonlinear response are now well established. However, under appropriate conditions, the nonlinear response of materials can be highly nonlocal, a phenomenon that drastically affects the propagation of intense laser radiation [2,3]. The nonlocality of nonlinear response comes into play when the transverse extent of the laser beam becomes comparable with the characteristic response length of the medium. The nonlocal nonlinear response allows suppression of the modulation instability of the plane waves in focusing media [4], it prevents catastrophic collapse of multidimensional beams [5,6], and stabilizes complex soliton structures, including vortex solitons [7]. Principally new effects attributed to nonlocality have been studied in photorefractive crystals [8], thermo-optical materials [9], liquid crystals [10], plasmas [11], and Bose-Einstein condensates with long-range interparticle interactions [12].

Soliton properties are also strongly altered by transverse modulations of refractive index. One of the principal properties featured by the corresponding discrete or lattice solitons, is their restricted mobility in the transverse plane, the effect which might be employed for various switching and routing operations [13,14]. Recent progress in creation of reconfigurable optical lattices in photorefractive crystals [15] and nematic liquid crystals [10] opened a direct way to explore the properties of solitons by varying the lattice depth and period. However, photorefractive media and liquid crystals may feature a strong nonlocal nonlinear response. Therefore, a principal question arises about the effect of the interplay of periodic refractive index modulation and nonlocality of nonlinear response on fundamental soliton properties, including their mobility. An intuitively similar, but physically drastically different scenario is the tunable self-bending of solitons in lattices made in media with diffusive nonlinearity [16]. In this Letter, we address the properties of solitons in Kerr-type *nonlocal* nonlinear media with an imprinted transverse periodic modulation of the refractive index. Our central discovery is that a *tunable nonlocality* can greatly enhance the soliton mobility, opening the possibility of almost radiationless soliton propagation across the lattice. We employ a generic model for the nonlocal nonlinearity, which provides insight for all physical settings governed by nonlocality kernels with an exponential-decaying range, including photorefractive and liquid crystal optical media, as well as in models of Bose-Einstein condensates with long-range interparticle interactions [12].

For concreteness, here we consider the propagation of the light beam along the $z$ axis in a nonlocal nonlinear Kerr-type medium with an imprinted modulation of linear refractive index described by the system of phenomenological equations for dimensionless complex light field amplitude $q$ and nonlinear correction to the refractive index $n$ [3-6]:

$$i\frac{\partial q}{\partial \xi} = -\frac{1}{2}\frac{\partial^2 q}{\partial \eta^2} - qn - pR(\eta)q,$$

$$n - d\frac{\partial^2 n}{\partial \eta^2} = |q|^2, \qquad (1)$$

where $\eta$ and $\xi$ stand for the transverse and longitudinal coordinates scaled to the beam width and diffraction length, respectively; the parameter $d$ stands for the degree of nonlocality of the nonlinear response; the parameter $p$ is proportional to the refractive index modulation depth; and the function $R(\eta) = \cos(2\pi\eta/T)$ describes the transverse refractive index profile, where $T$ is the modulation period. We assume that the depth of the refractive index modulation is small compared to the unperturbed index. In the limit $d \to 0$, the system (1) reduces to the nonlinear Schrödinger equation. The opposite case $d \to \infty$ corresponds to the strongly nonlocal regime. Among the conserved quantities of system (1) are the energy flow $U$ and the Hamiltonian $H$

$$U = \int_{-\infty}^{\infty} |q|^2 \, d\eta,$$

$$H = \int_{-\infty}^{\infty} \left[ \frac{1}{2}\left|\frac{\partial q}{\partial \eta}\right|^2 - pR(\eta)|q|^2 - \frac{1}{2}|q|^2 \int_{-\infty}^{\infty} G(\eta - \lambda)|q(\lambda)|^2 \, d\lambda \right] d\eta, \qquad (2)$$

where $G(\eta) = (1/2d^{1/2})\exp(-|\eta|/d^{1/2})$ is the response function of the nonlocal medium. We search for stationary solutions of Eq. (1) in the form $q(\eta,\xi) = w(\eta)\exp(ib\xi)$, where $w(\eta)$ is a real function and $b$ is a real propagation constant. Substitution into (1) yields

$$\frac{d^2w}{d\eta^2} + 2wn + 2pRw - 2bw = 0,$$
$$d\frac{d^2n}{d\eta^2} - n + w^2 = 0,$$
(3)

where $n$ stands for the stationary refractive index profile. We solved these equations numerically with a relaxation method. We set $T = \pi/2$ and vary $b$, $p$, and $d$. To elucidate the linear stability of the solitons, we searched for perturbed solutions in the form $q(\eta,\xi) = [w(\eta) + u(\eta,\xi) + iv(\eta,\xi)]\exp(ib\xi)$, where the real $u(\eta,\xi)$ and imaginary $v(\eta,\xi)$ parts of the perturbation can grow with a complex rate $\delta$ upon propagation. Linearization of Eq. (1) around a stationary solution yields the eigenvalue problem

$$\delta u = -\frac{1}{2}\frac{d^2v}{d\eta^2} + bv - nv - pRv,$$
$$\delta v = \frac{1}{2}\frac{d^2u}{d\eta^2} - bu + nu + w\Delta n + pRu,$$
(4)

where $\Delta n = 2\int_{-\infty}^{\infty} G(\eta - \lambda)w(\lambda)u(\lambda)d\lambda$. The system (4) can also be solved numerically.

First we address properties of lowest-order odd and even solitons. The absolute intensity maximum for odd solitons coincides with one of the local maxima of $R(\eta)$ (Fig. 1(a)), whereas even solitons are centered between neighboring lattice sites (Fig. 1(b)), and can be viewed as a nonlinear superposition of in-phase odd solitons. With increase of lattice depth the soliton energy concentrates in the guiding lattice sites (the regions of local refractive index maxima) so that lattice solitons approach their discrete counterparts [13]. The energy flow $U$ for both odd and even solitons is a monotonically growing function of the propagation constant $b$ and it vanishes in the cutoff $b_{co}$ point, which is identical for odd and for even solitons (Fig. 1(c)). The cutoff $b_{co}$ for odd and even solitons is a monotonically growing function of lattice depth $p$ and we found that it *does not depend* on the nonlocality degree $d$. This is the consequence of the fact that both odd and even solitons reside in the semi-infinite gap of Floquet-Bloch spectrum of linear lattice that is independent of the nonlocality degree $d$ so that $b_{co}$ always coincides with the lower edge of this gap (see Ref. [23] for detailed discussion of band-gap lattice structure and bifurcations of gap solitons in local cubic media). At fixed energy flow and lattice parameters soliton gets broader and its peak amplitude decreases with increase of $d$. Linear stability analysis revealed that odd solitons are stable and even solitons are unstable in the entire domain of their existence, similarly to the case of local medium [23]. However, the perturbation growth rate for even soliton is drastically reduced with increase of nonlocality (Fig. 1(d)), so that even solitons with moderate energy flows $U$ can propagate undistorted in highly nonlocal medium, even in the presence of random perturbations of the initial conditions, over distances exceeding any experimentally feasible crystal length by several orders of magnitude. Therefore, a first important result uncovered is that *nonlocality largely reduces* the strength of symmetry-breaking instabilities.

We also found families of twisted solitons that can be considered combinations of several odd solitons with engineered phases [17] (see Fig. 2 for illustrative examples). The energy flow of twisted solitons is a nonmonotonic function of propagation constant and there exist a lower cutoff for existence of such solitons (Fig. 2(c)). The slope $dU/db$ of the curve $U(b)$ becomes negative in a narrow region near the cutoff, not even visible in Fig. 2(c). Contrary to the case of odd and even solitons the cutoff $b_{co}$ for twisted solitons increases with increase of the nonlocality degree $d$. Stability analysis revealed that twisted solitons feature both exponential and oscillatory instabilities near the lower cutoff for their existence (see Fig. 2(d)). However, we found that they become completely stable above a certain energy flow threshold. The width of the instability domain for twisted solitons decreases with increase of lattice depth and increases with growth of nonlocality degree.

As one can see from panels (a) and (b) of Figs. 1 and 2, for all solitons found the nonlinear refractive index distribution in nonlocal media with $d \sim 1$ always features smooth symmetric bell-like shape without pronounced local maximums on top of it thereby smoothing over the total refractive index profile $n + pR$. This is in clear contrast to local cubic medium, where focusing nonlinearity tends to further increment the transverse refractive index modulation that, in turn, results in a restricted mobility of high-energy excitations. Therefore, the nonlocality of nonlinear response could greatly enhance transverse soliton mobility, which is the central result of this Letter.

Figure 3 confirms this central result. The plot shows the Peierls-Nabarro (PN) barrier, defined as a difference $\delta H = H_{even} - H_{odd}$ between Hamiltonians for even and odd solitons carrying the same energy flow $U$ [18]. Since upon motion across the lattice the solitons pass through odd and even states, thus accompanied by the corresponding changes in Hamiltonian (or potential energy, when solitons are viewed as particles), the higher the barriers the larger the incident angle (or kinetic energy) required to overcome them. As expected, the height of the PN barrier grows with increase of soliton energy flow $U$ and lattice depth $p$ (Fig. 3(a)). However, the *nonlocality reduces drastically* the value of PN barrier (Fig. 3(b)). The physical implication is that corresponding solitons can move across the lattice almost without radiation losses, because even small angles are sufficient to overcome the reduced PN barrier. Therefore an increasing degree of nonlocality affords very significant enhancement of the mobility of high-energy lattice solitons, a feature with both fundamental and potential practical relevance. Since odd solitons are ground-state solutions and realize the most energetically favorable state of the system, the difference $\delta H$ can also serve as a measure of the instability of even solitons that is drastically reduced with increase of nonlocality degree. Notice that on physical grounds, the enhanced mobility of nonlocal lattice solitons cannot be attributed to any variation of soliton stability (as it occurs in some discrete systems with intersite interactions [19]), but solely to refractive index smoothing induced by the nonlocality.

The expectations based on the reduction of the PN barrier are fully confirmed by numerical integration of Eq. (1). Fig. 4 illustrates the point. To stress the physical robustness of our findings, here we present results obtained with inputs in the form $q(\eta, \xi = 0) = \chi \operatorname{sech}(\chi \eta) \exp(i\alpha \eta)$, where $\alpha$ stands for the incident angle. In general, soliton motion across the lattice is accompanied by radiative losses that eventually lead to soliton capture in one of the lattice channels (Fig. 4(c)). The radiation losses are drastically reduced by the nonlocality. For example, a soliton with $\chi = 1.2$ trapped in $10^{\text{th}}$ channel of the lattice imprinted in fully local medium looses about 40% of its input energy flow, while in nonlocal medium with $d = 0.3$ the energy losses are less than 10%. Further decreases of the radiation are achieved by increasing the degree of nonlocality. Let the soliton be trapped in the *N*-th channel if $NT - T/2 < \eta_{\max} < NT + T/2$ at $\xi \to \infty$, where $\eta_{\max}$ is the transverse coordinate of soliton center. The output channel number decreases with increase of lattice depth $p$ (Fig. 4(a)), but we found that it *does grow with increase of nonlocality* (Fig. 4(b)). Importantly, small variations of the nonlocality impact strongly the value of the output channel number, a result that stresses the new degree of freedom afforded by the nonlocality.

It is worth stressing that the physical mechanism behind the enhanced soliton mobility in nonlocal media put forward here is principally different from the tunable bending that occurs in media with *diffusion asymmetric nonlinearity* [16], where highly anisotropic nonlinear response results in asymmetric self-induced refractive index profiles and causes soliton bending and lattices are used to tune the bending rate. In contrast, the principal physical feature behind the phenomena uncovered here is the *tunability of a symmetric nonlocality*, itself. Such aim can be achieved in a variety of ways. In particular it has been experimentally demonstrated that the nonlocality of nematic liquid crystals vary with the voltage applied the crystal [20]. Increase in the voltage causes reorientation of molecules of liquid crystal, which, in turn, results in modification of character of nonlinear response from highly nonlocal to predominantly local. Since lattices can be imprinted in liquid crystals [10], they are very promising candidates for demonstration of enhanced mobility of nonlocal lattice solitons. Thus, the variation of the output channel upon slight modification of nonlocality degree depicted in Fig. 4 can be used to implement soliton-based switching and routing schemes controlled by the applied voltage. Another example of tunable nonlocality is encountered with thermal nonlinearity, e.g., in dye-doped liquid crystals [21].

In summary, we have addressed the properties of solitons propagating in optical lattices imprinted in Kerr-type nonlinear, nonlocal media. We revealed that the nonlocality introduces principal new effects into the soliton transverse mobility. In particular, we discovered that the Peierls-Nabarro potential barrier for solitons moving across the lattice is drastically reduced in the presence of the nonlocality, a result of fundamental importance because mobile lattice solitons appear to be very rare in nature [22]. Our predictions can be directly tested with light beams propagating in photorefractive and in liquid crystals, but we addressed a canonical, generic nonlocal model which provides insight for all analogous physical settings governed by symmetric nonlocality kernels featuring exponentially-decaying ranges.

*On leave from Physics Department, M. V. Lomonosov Moscow State University, Russia. This work has been partially supported by the Generalitat de Catalunya, by the Spanish Government through grant BFM2002-2861 and by the Ramon-y-Cajal program.

# Figure captions

Figure 1. Profile of (a) odd and (b) even solitons with energy flow $U = 4$ and corresponding nonlinear refractive index distributions. (c) Energy flow versus propagation constant for odd and even solitons. In (a), (b), and (c) degree of nonlocality $d = 2$. (d) Perturbation growth rate versus energy flow of even solitons at different degree of nonlocality. Lattice depth $p = 3$. Gray regions in (a) and (b) correspond to $R(\eta) \geq 0$, while in white regions $R(\eta) < 0$.

Figure 2. Profile of (a) first and (b) second twisted solitons and nonlinear refractive index distributions corresponding to points marked by circles in dispersion diagram (c). Lattice depth $p = 3$, nonlocality degree $d = 2$. (d) Real part of perturbation growth rate for first twisted soliton at $p = 2.5$ and various $d$ values.

Figure 3. Height of the PN barrier versus soliton energy flow at $d = 4$ (a) and versus degree of nonlocality at $p = 3$ (b).

Figure 4. (a) Output channel number versus lattice depth at $d = 0.08$. (b) Output channel number versus nonlocality degree at $p = 1$. (c) Soliton propagation trajectories at $d = 0.1$ (1), 0.26 (2), and 0.4 (3). Lattice depth $p = 1$. In all cases the input soliton form-factor $\chi = 1.2$ and incident angle $\alpha = 0.5$.

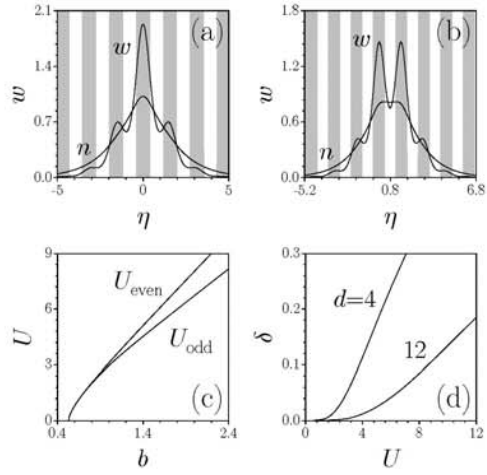

Figure 1. Profiles of (a) odd and (b) even solitons with energy flow $U = 4$ and corresponding nonlinear refractive index distributions. (c) Energy flow versus propagation constant for odd and even solitons. In (a), (b), and (c) degree of nonlocality $d = 2$. (d) Perturbation growth rate versus energy flow of even soliton at different degrees of nonlocality. Lattice depth $p = 3$. Gray regions in (a) and (b) correspond to $R(\eta) \geq 0$, while in white regions $R(\eta) < 0$.

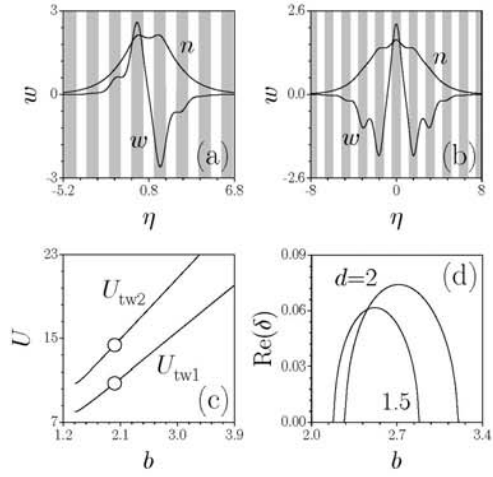

Figure 2. Profiles of (a) first and (b) second twisted solitons and nonlinear refractive index distributions corresponding to points marked by circles in dispersion diagram (c). Lattice depth $p=3$, nonlocality degree $d=2$. (d) Real part of perturbation growth rate for first twisted soliton at $p=2.5$ and various $d$ values.

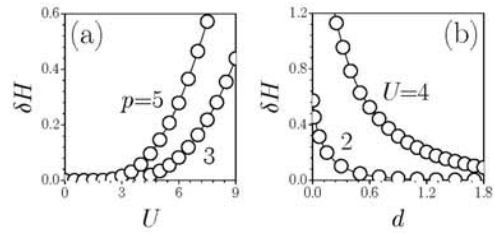

Figure 3.  The height of PN barrier versus soliton energy flow at $d=4$ (a) and versus nonlocality degree at $p=3$ (b).

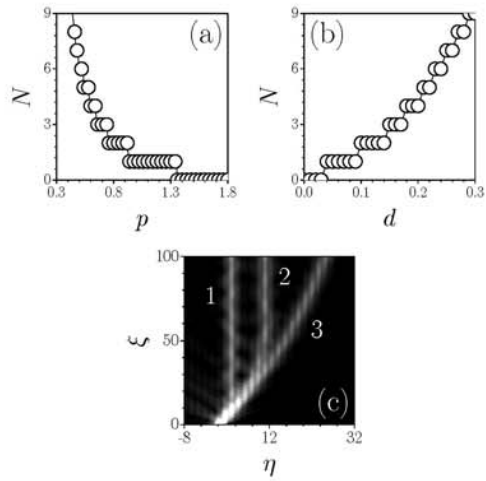

Figure 4. (a) Output channel number versus lattice depth at $d = 0.08$. (b) Output channel number versus nonlocality degree at $p = 1$. (c) Soliton propagation trajectories at $d = 0.1$ (1), $0.26$ (2), and $0.4$ (3). Lattice depth $p = 1$. In all cases the input soliton form-factor $\chi = 1.2$ and incident angle $\alpha = 0.5$.